\documentclass{article}
\usepackage{epsfig}
\usepackage{amssymb}
\usepackage{amsmath}
\usepackage{amsfonts}
\newcommand \be{\begin{eqnarray}}
\newcommand \ee{\end{eqnarray}}
\begin{document}
\begin{center}
{\bf Spin 1/2 Fermions in the Unitary Limit.II}\\
\bigskip
\bigskip
H. S. K\"ohler \footnote{e-mail: kohler@physics.arizona.edu} \\
{\em Physics Department, University of Arizona, Tucson, Arizona
85721,USA}\\
\end{center}
\date{\today}

\begin{abstract}
This report concerns the energy of a zero-temperature many-body system of 
spin $\frac{1}{2}$ fermions in the unitary limit. In a
previous report (nucl-th/0705.0944) this energy was determined to be $\xi=0.24$
in units of the free gas kinetic energy, appreciably lower than
most reports  giving $\xi\sim 0.45$. In our calculation the 2-body
interaction satisfied exactly the unitary limit i.e. 
infinite scattering length and effective range $r_0=0$. In the present report 
results with $r_0>0$ are shown. A strong dependence on the effective
range is found. It is  for example found that
an increase to $r_0=1fm$ increases  $\xi$ to $\sim 0.4$ close to other
reports of $\xi$ in the unitary limit. It is concluded that 
because of the singular character of
the unitary limit it is necessary to verify that the interaction
actually satisfies unitarity.
 
The calculations done here in a pp-ladder approximation show a
resonance in the \it in-medium \rm interaction close to (and in) the unitary
limit. This was already found in the previous work.
\end{abstract}
 
\section {Introduction}
The properties of  a dilute fermigas with large scattering length 
is of considerable theoretical as well as experimental interest. 

Several numerical methods have been used to determine $\xi$
with results varying from $\xi\sim 0.5$ to $\xi\sim 0.25$. Extensive
references are for example found in \cite{bul07,siu07}
 
Apart from the interest in the properties of a unitary gas these
calculations also provide a test of many-body methods.  The Monte
Carlo calculations are at least in principle the most accurate and
could provide a benchmark.

It is a well-known fact that interactions with  large scattering lengths
are separable.
This paper is a  report on results of  calculations
using  separable interactions 
determined by inverse scattering from phase-shifts of various large
scattering lengths $a_s$ and small effective ranges including $a_s=\infty$
and $r_0=.0$.
The energy at  zero temperature is calculated
from an in-medium effective interaction obtained from a pp-ladder
summation of these separable interactions. 

Section 2 is a short summary of the method used with
numerical results  shown in Section 3
and some of the conclusions are summarised in Section 4.

\section{Separable Interaction from inverse scattering}
The method used here to calculate a separable interatction by inverse
scattering has also been
used in several previous papers where details can be
found.\cite{kwo95,hsk07,hsk04,hskm07}
Only a short summary of the expressions is given here.
The input in the calculations are phase-shifts, either experimental or
otherwise defined as shown below. 
A rank one separable potential provides a sufficient and in fact precise
description of the interaction in the unitary limit. If the phase-shift
changes sign such as in the experimental $^1S_0$ case a rank two potential
is necessary. (see ref \cite{kwo95}).
In the case of a rank one attractive potential one has

\begin{equation}
V(k,p)=-v(k)v(p)
\label{V}
\end{equation}
Inverse scattering then yields
(e.g. ref \cite{kwo95,tab69})

\begin{equation}
v^{2}(k)= \frac{(4\pi)^{2}}{k}sin \delta (k)|D(k^{2})|
\label{v2}
\end{equation}
where
\begin{equation}
D(k^{2})=exp\left[\frac{2}{\pi}{\cal P}\int_{0}^{\Lambda}
\frac{k'\delta(k')}{k^{2}-k'^{2}}dk' \right]
\label{D}
\end{equation}
where ${\cal P}$ denotes the principal value  and
$\delta(k)$ is the phaseshift. $\Lambda$ provides a cut-off in
momentum-space. 
The effect of the  cut-off will be exploited below.

With $\delta(k)=\pi/2$, the unitary limit,  one finds 
\begin{equation}
v_u^{2}(k)= -\frac{(4\pi)^{2}}{(\Lambda^{2}-k^{2})^{\frac{1}{2}}}
\label{vpi2}
\end{equation}
and $v_u^2(k)$
$\rightarrow -\infty$ for $k\rightarrow \Lambda$.(See Fig \ref{unipot}).

For $\Lambda \gg k$ one finds
\begin{equation}
v_u^{2}(k)\rightarrow -\frac{(4\pi)^{2}}{\Lambda}
\label{v2ka}
\end{equation}
In this limit, \it but only in this limit \rm , 
the unitary interaction will then be a
$\delta$-function in coordinate space. And the strength is inversely
proportional to the cut-off.

The $V_{low \ k}$ approximation is  adequate at low density 
for some of the cases shown below but NOT in the unitary limit. 
An effective in-medium Brueckner interaction as defined by a
particle-particle ladder summation is used for all the results 
presented here. 
Dispersion corrections are expected to
be small \cite{hskm07} so that the denominator  has only kinetic energies;
the effective mass $m^*=1$.
The diagonal elements of the in-medium interaction is then

\begin{equation}
 G(k,P)=-\frac{v^{2}(k)}{1+ I_{G}(k,P)}
\label{G}
\end{equation}
with
\begin{equation}
I_{G}(k,P)=\frac{1}{(2\pi)^{3}}\int_{0}^{\Lambda}
v^{2}(k')\frac{Q(k',P)}{k^{2}-k'^{2}}
k'^{2}dk'
\label{I_G}
\end{equation}
where $P$ is the center of mass momentum and  $Q$ the angle-averaged
Pauli-operator for pp-ladders (Brueckner approximation). 
One should note that 
the angle-averaging  is exact here because the denominator is independent
of $P$. 

It was shown in ref \cite{hsk07} that this can be rewritten as

\begin{equation}
 G(k,P)=-\frac{v^{2}(k)}{I_{GK}(k,P)}
\label{GK}
\end{equation}
with
\begin{equation}
I_{GK}(k,P)=\frac{1}{(2\pi)^{3}}\int_{0}^{2k_f}
v^{2}(k')\frac{Q(k',P)-{\cal P}}{k^{2}-k'^{2}}
k'^{2}dk'+\frac{kv^{2}(k)}{\tan\delta(k)}
\label{IGK}
\end{equation}

Eqs (\ref{GK},  \ref{IGK}) have the advantage over eqs (\ref{G}, \ref{I_G}) 
in that the
integrand in eq (\ref{IGK})
is zero for  $k'>2k_f$
because   the 
factor $Q(k',P)-\cal{P}$ is then equal to zero. 
Consequently, there is no need to resort to a low-momentum ("$V_{low \ k}$")
approximation here. Although the two sets of equations are numerically
identical the latter set simplifies the computing greatly. This is in
particular important in the unitary limit in which case the integration can
be done analytically to give eq (\ref{Ipp}).

With $a=\frac{k}{k_{f}}$ and $y=\frac{P}{2k_{f}}$
the potential energy per particle $PE/A$ is
\begin{equation}
PE/A=\frac{3k_{f}^{3}}{\pi^{2}}\int_{0}^{1}\left[\int_{0}^{1-a}8G(a,y)y^{2}dy+
\frac{1}{a}\int_{1-a}^{(1-a^{2})^{\frac{1}{2}}}4G(a,y)(1-y^{2}-a^{2})ydy\right]
a^{2}da
\label{PE}
\end{equation}

The kinetic energy per particle, i.e. the uncorrelated fermi-gas energy
is given by $$E_{FG}/A=\frac{3}{10}\frac{\hbar^{2}}{m}k_{f}^{2}.$$
The total energy is expressed in these units by
$$E/A=\xi E_{FG}/A.$$

In the unitary limit ($\delta(k)=\pi/2$) there is a simplification already
used in ref \cite{hsk07}. The cutoff $\Lambda$ can then be chosen large so that eq
(\ref{v2ka}) is valid and with $v_u(k')$ independent of $k'$ 
one can perform the $k'$-integration in eq (\ref{IGK}) analytically. 
After dividing by $v_u^2$ one then finds

\begin{equation}
I_u(a,y)=
\frac{k_{f}}{\pi}\left[1+y+a*log\left|\frac{1+y-a}{1+y+a}\right|
+\frac{1}{2y}(1-y^{2}-a^{2})log\left|\frac{(1+y)^{2}-a^{2}}{1-y^{2}-a^{2}}
\right|\right].
\label{Ipp}
\end{equation}
and
\begin{equation}
G(a,y)=-4\pi[I_u(a,y)]^{-1}
\label{Gay}
\end{equation}

This then provides an analytic expression for $G(a,y)$ and as already shown in
ref \cite{hsk07} one finds in this case $\xi=0.24$, explicitly independent of 
$\Lambda$ and of density (fermi-momentum) as is to be expected in the 
unitary limit. 

It is of interest that Steele \cite{ste00}, using the 
effective field theory power counting method arrives at the
same expression for $G(a,y)$. But in eq \ref{PE} for the potential energy he
uses the approximation $G(a,y)\sim G(0,0)$ and then arrives at $\xi=4/9$.

\section{Numerical Results}
The expression (\ref{IGK}) was  used in all calculations together with the
analytic expression (\ref{Ipp}), valid the in unitary limit. 

Calculations were made with phase-shifts defined as follows:\\
\\
A: The experimental $^1S_0$ phases.\\
B: Scattering length $a_s=-18.5fm$ and effective range $r_0=2.8fm$.\\
C: Scattering length $a_s\rightarrow-\infty$ and effective range $r_0=2.8fm$.\\
D: Scattering length $a_s\rightarrow-\infty$ and effective range $r_0=1.0fm$.\\
E: Scattering length $a_s\rightarrow-\infty$ and effective range $r_0=0.5fm$.\\
F: Scattering length $a_s\rightarrow-\infty$ and effective range $r_0=0.fm$ and
   the interaction given by eq. (\ref{vpi2}).\\
G: Scattering length $a_s\rightarrow-\infty$ and effective range $r_0=0.fm$ and
   the interaction given by eq. (\ref{v2ka}).\\

\begin{figure}
\centerline{
\psfig{figure=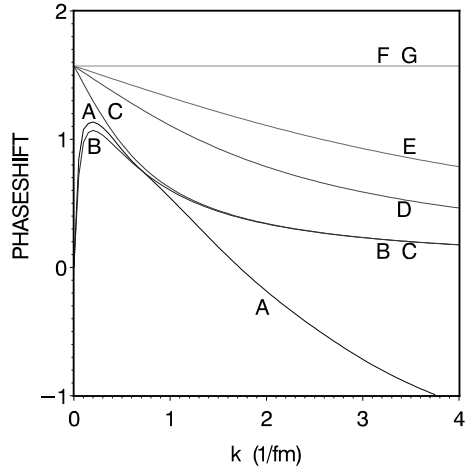,width=7cm,angle=0}
}
\vspace{.0in}
\caption{This figure shows the phase-shifts that were used as input for the
calculations with the labels defined in the text below.}
\label{uniphase}
\end{figure}

Fig \ref{uniphase} shows these phase-shifts as a function of momentum
for $k<\Lambda=4 fm$. Notice that phases "A" and "B" are comparable for
small momenta, that "B" and "C" overlap except for
small momenta and that "F" and "G" are for the unitary limit,
$\delta=\frac{\pi}{2}$.

\begin{figure}
\centerline{
\psfig{figure=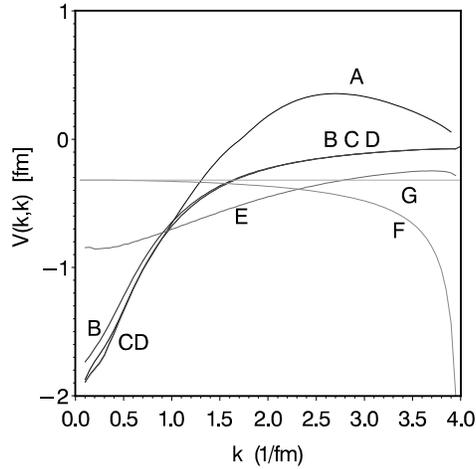,width=7cm,angle=0}
}
\vspace{.0in}
\caption{This figure shows the diagonal in momentum-space 
of the potentials corresponding to the phase-shifts in Fig \ref{uniphase}.}
\label{unipot}
\end{figure}

The potentials correponding to each of these sets of phase-shifts were
calculated using the inverse scattering eqs (\ref{v2}) and (\ref{D}) with the 
results shown in Fig \ref{unipot}. The potential curves A
$\rightarrow$ D
with  effective ranges $r_0\leq 1.0$ 
are (practically) overlapping for $k<1 fm^{-1}$ 
even though the scattering length varies over a large interval,
$-\infty<a_s<-18.5$. Note however the drastic change of the potentials when
$r_0$ changes from $1$ to $0.5$ to $0$ (curves labelled D,E and F) while
$a_s\rightarrow -\infty$. Note that curve "F" being the exact solution for a
separable potential in the unitary limit is uniquely different from all the
others.

\begin{figure}
\centerline{
\psfig{figure=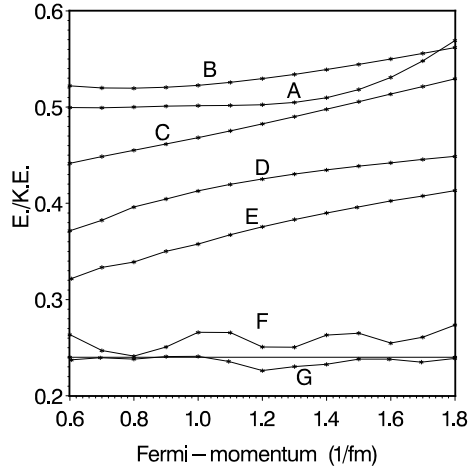,width=7cm,angle=0}
}
\vspace{.0in}
\caption{This figure shows the energy (in units of the fermi-gas energy)
for various fermi-momenta. The somewhat "wavy" appearance of the curves "F"
and "G" (the unitary limit) is because of numerical difficulties described
in the text. The straight line is the explicitly density-independent result
from an analytic solution of the effective interaction $G$ in the unitary
limit.\cite{hsk07}}
\label{unixi}
\end{figure}

The total energy $\xi$ in units of the fermigas-energy
$E_{FG}/A$ is shown in Fig \ref{unixi}. Our curve "A" for the $^1S_0$ phases
agrees practically exactly with the corresponding curve of ref \cite{siu07}
even though the latter also includes some ring diagrams. Curve "B" with the
"$^1S_0$" scattering length and effective range gives a somewhat 
larger value of $\xi$.
Note that, with $a_s$ held constant at
$-\infty$, $\xi$ changes from $\sim 0.4\rightarrow 0.5$ for  $r_0=2.8$ 
to $\sim 0.24$ for $r_0=0$. 
The value $\xi=0.24$ independent of $k_f$ is indicated by the straight
line. It was already obtained in  ref \cite{hsk07} using eq (\ref{Ipp}).
Curve "F" and more so, curve "G", nearly coincide with this analytically
obtained result. Curve "G" uses the same large $\Lambda$ limit of $v_u$
given by eq (\ref{v2ka}) as in the analytic calculation, while "F" is
calculated with $\Lambda=6 fm^{-1}$.
The value of $\xi=0.24$, given by the straight line, is the unitary limit
result from the analytic calculation. It is only in this limit that the
analytic integration is possible. The numercal results represented by
curves "F" and "G" are necessarily less accurate as discussed below.
Those calculations were done only 
in order to confirm computational consistency.

It was already pointed out in ref \cite{hsk07} that, in the unitary
limit,
$I_{GK}=0$ along a line in the $(a,y)$-plane from $\sim (0.8,0)$ to
$\sim(0.9,0.3)$. The associated singularity in $G(a,y)$
implies a resonance with the \it in-medium \rm
phase-shift $\rightarrow \pi/2$ which appears very unusual. 
It may however not 
survive in a higher order calculation, in particular including spectral
broadening as in Green's function calculations. 
The singularity in $G(a,y)$ complicates the numerical integration in eq
(\ref{PE}) and an 
interpolation method was used in ref \cite{hsk07}. The same method was now used
for obtaining the "F" and "G" results for $\xi$. Instead of the analytic
expression (\ref{Ipp}) the numerically integrated eq (\ref{IGK}) has now to
be interpolated and this results in less accuracy resulting in the
oscillations seen in Fig \ref{unixi} especially for curve "F".
Because the analytic solution is more accurate it supersedes the
"F" and "G" and an improvement in
the numerical calculation to remedy this situation was ruled unnecessary.

\section{Conclusions}
It is , we believe, conclusively shown that a serious study of a system in the
unitary limit has to be done in the limit itself. Any extrapolation is
questionable, especially as regards the effective range. 
The unitary limit is (in a broad sense) singular and unique. This is the
necessary conclusion from Fig \ref{unixi}.
It is in fact suggested already by studying the
phase-shifts shown in Fig \ref{uniphase} and even more so the
potential-curves in Fig \ref{unipot}. 
There is no reason to
believe that this conclusion is unique to the methods used here.
The strong
dependence of $\xi$ on the effective range $r_0$ is an important
effect and the major result of this investigation.

Using eqs (\ref{GK}) and (\ref{IGK}) the cut-off can be chosen
arbitrarly large, $\Lambda\gg 2k_f$ and eq (\ref{v2ka}) is then a valid
approximation in the unitary limit.  
This presents a great simplification resulting in
the analytic expression (\ref{Gay}) for the effective interaction. This same
result was obtained by Steele\cite{ste00} using power-counting. 
Eqs (\ref{G},\ref{I_G}) are numerically
equivalent with (\ref{GK},\ref{IGK}) but requires 
using eq (\ref{vpi2}) for the interaction resulting
in a much more difficult integration with the upper limit equal to
$\Lambda$ where the unitary interaction (\ref{vpi2}) diverges.

As for the numerical value of $\xi$ it is not claimed to be decided by 
the zero-temperature ladder approximation used here.
Dispersion (mean field) corrections wre not included as
only kinetic energies were used in the propagators, i.e. $m^*=1$. This
approximation led to eq (\ref{IGK}) and eq (\ref{Ipp}) which provides a
great simplification and indeed made the calculations feasible.
The dispersion correction is small for
the $^1S_0$ state \cite{hskm07} and although not proven it is 
expected to be small also in the unitary limit. 
Dispersion-corrections are related not only to the mean field but also to
the in-medium two-body corrrelations.\cite{hskm07}. It is in general
repulsive and would therefore increase the value of $\xi$. It can not be
ruled out that it is larger than expected.

Pairing corrections are important. The critical temperature
is relatively high (e.g. \cite{bul07}) and the boson-fraction is high at
zero temperature.

It is generally accepted that Monte Carlo calculations is the best
approach to solving the unitary problem. The present investigation is
presented only to suggset the importance of using an excatly "unitary
interaction".
   
The calculations were based on using separable interactions obtained by
inverse scattering from  phase-shifts. Earlier work has shown 
the near equivalence of this method with the conventional potential
approach\cite{kwo95}. It is therefore expected that the conclusions 
obtained here are
not unique to our use of these separable potentials but are in fact general.

In a future publication a report with Green's function techniques allowing
a finite temperature calculation including hh-ladders will be reported.
These calculations are more extensive than the ladder summations used here.
The effects of spectral broadening as well dispersion effects are included. 

As already pointed out above
there are some similarities of these calculations with those of ref
\cite{siu07} although that work only considers
$r_0\geq 2.54fm$ for which there is however a semi-quantative agreement between 
the respective results even though  a 
type of ring-diagrams are included in ref \cite{siu07} and the interactions
are also different.

After this report was first submitted to the arxiv the author became aware
of a work by T. Sch\"aefer et al \cite{sch05}
extending Steele's work. One result of
their work was the same value $\xi=0.24$ referred to above and reported in
ref.  \cite{hsk07}. They also obtain a strong dependence on effective range
as reported here.

I thank Professor Sch\"aefer for kindly bringing this publication to my
attention.

\end{document}